\documentclass[8.5pt,twoside,twocolumn]{article}
\usepackage[super,sort&compress,comma]{natbib}
\usepackage{mhchem}
\usepackage{times,mathptmx}
\usepackage{balance}

\usepackage{graphicx} 
\usepackage[format=plain,justification=raggedright,singlelinecheck=false,font=small,labelfont=bf,labelsep=space]{caption}
\usepackage{fancyhdr}
\begin{document}

\thispagestyle{plain}
\fancypagestyle{plain}{
\renewcommand{\headrulewidth}{1pt}}
\renewcommand{\thefootnote}{\fnsymbol{footnote}}
\renewcommand\footnoterule{\vspace*{1pt}%
\hrule width 3.4in height 0.4pt \vspace*{5pt}}
\setcounter{secnumdepth}{5}

\makeatletter
\def\subsubsection{\@startsection{subsubsection}{3}{10pt}{-1.25ex plus -1ex minus -.1ex}{0ex plus 0ex}{\normalsize\bf}}
\def\paragraph{\@startsection{paragraph}{4}{10pt}{-1.25ex plus -1ex minus -.1ex}{0ex plus 0ex}{\normalsize\textit}}
\renewcommand\@biblabel[1]{#1}
\renewcommand\@makefntext[1]%
{\noindent\makebox[0pt][r]{\@thefnmark\,}#1}
\makeatother
\renewcommand{\figurename}{\small{Fig.}~}

\fancyfoot{}
\fancyfoot[RO]{\footnotesize{\sffamily{1--\pageref{LastPage} ~\textbar  \hspace{2pt}\thepage}}}
\fancyfoot[LE]{\footnotesize{\sffamily{\thepage~\textbar\hspace{3.45cm} 1--\pageref{LastPage}}}}
\fancyhead{}
\renewcommand{\headrulewidth}{1pt}
\renewcommand{\footrulewidth}{1pt}
\setlength{\arrayrulewidth}{1pt}
\setlength{\columnsep}{6.5mm}
\setlength\bibsep{1pt}

\twocolumn[
  \begin{@twocolumnfalse}
\noindent\LARGE{\textbf{Formation of ultracold metastable RbCs molecules by short-range photoassociation}}
\vspace{0.6cm}

\noindent\large{\textbf{C.Gabbanini$^{\ast}$\textit{$^{a}$} and O.Dulieu\textit{$^{b}$}}}\vspace{0.5cm}

\noindent\textit{\small{\textbf{Received Xth XXXXXXXXXX 20XX, Accepted Xth XXXXXXXXX 20XX\newline
First published on the web Xth XXXXXXXXXX 200X}}}

\noindent \textbf{\small{DOI: 10.1039/b000000x}}
\vspace{0.6cm}

\noindent \normalsize{Ultracold metastable RbCs molecules are observed in a double species MOT through photoassociation near the Rb(5S$_{1/2}$)+Cs(6P$_{3/2}$) dissociation limit followed by radiative stabilization. The molecules are formed in their lowest triplet electronic state and are detected by resonant enhanced two-photon ionization through the previously unobserved $(3)^{3}\Pi \leftarrow a^{3}\Sigma^{+}$ band. The large rotational structure of the observed photoassociation lines is assigned to the lowest vibrational levels of the $0^+,0^-$ excited states correlated to the Rb(5P$_{1/2}$)+Cs(6S$_{1/2}$) dissociation limit. This demonstrates the possibility to induce direct photoassociation in heteronuclear alkali-metal molecules at short internuclear distance, as pointed out in [J. Deiglmayr \textit{et al.}, Phys. Rev. Lett. \textbf{101}, 13304 (2008)].}
\vspace{0.5cm}
 \end{@twocolumnfalse}
  ]

\section{Introduction}
\footnotetext{}
The field of cold molecules received a large attention in the last decade due to important advances and potentially new applications in several domains like fundamental tests in physics, molecular clocks, molecular spectroscopy, dynamics of cold reactions, controlled photochemistry studies and quantum computation~\cite{Fried2009,dulieu2009,carr2009}. In particular cold polar molecules (i.e. exhibiting a permanent electric dipole moment) thanks to long range dipolar interactions have been proposed for quantum information~\cite{demille2002,rabl2007}. The large anisotropic interaction between cold polar molecules is expected to give rise to quantum magnetism \cite{barnett2006} and to novel quantum phases \cite{baranov2002}. Polar molecules can be manipulated by external electric fields, allowing for the control of elementary chemical reactions at very low temperatures \cite{krems2005}.

The techniques that have produced up to now molecules in the ultracold temperature range are magnetoassociation and photoassociation (PA). PA of laser-cooled atoms followed by stabilization via spontaneous emission has been very successful for many homonuclear and heteronuclear alkali-metal molecules~\cite{dulieu2009}. The molecules are typically formed in high vibrational levels of the ground singlet or triplet state. The population can be transferred to lower vibrational levels, eventually to the $v=$0 level, by incoherent~\cite{sage2005} or coherent optical processes. In particular the stimulated Raman adiabatic passage (STIRAP) process has demonstrated to be a powerful tool for quantum degenerate gases starting from molecules produced by magnetoassociation~\cite{lang2008a,ni2008,danzl2008}, and recently also for ultracold molecules created by PA in a magneto-optical trap~\cite{aikawa2010}. Another possible approach is to perform vibrational cooling by a properly shaped laser~\cite{viteau2008}. For both methods a good spectroscopic knowledge of the molecule under study and a favourable rovibrational population distribution are crucial, therefore it is important to investigate different formation paths.

For what concerns ultracold RbCs molecules, their formation in the ground $a^{3}\Sigma^{+}$ state has been reported by PA below the lowest excited asymptote Rb($5S_{1/2}$)+Cs($6P_{1/2}$) and subsequent spontaneous emission~\cite{kerman2004a}. $^{85}$Rb and Cs atoms were excited into the $0^{-}$ state resulting from the coupled $(2) ^{3}\Sigma^{+}$ and $b ^{3}\Pi$ states with the decay process producing RbCs in the ground triplet state ~\cite{kerman2004}.  The molecules were detected by state selective REMPI through the coupled $(2)^{3}\Sigma^{+}$ and $B^{1}\Pi$ states. Some RbCs molecules were transferred in a following experiment to the $v=$0 level of the $X^{1}\Sigma^{+}$ ground state by an incoherent optical pumping process~\cite{sage2005}. The process was based on the $c ^{3}\Sigma^{+}$, $B ^{1}\Pi$ and $b ^{3}\Pi$ coupled states, allowing to circumvent the triplet-singlet forbidden transition, using two lasers for pumping and dumping the molecules from the starting level ($v'$=37 level of the $a ^{3}\Sigma^{+}$ state).

In this paper a different formation path of triplet RbCs molecules is reported, that involves PA near the Rb(5S$_{1/2}$)+Cs(6P$_{3/2}$) dissociation limit followed by radiative stabilization. The produced molecules are detected by resonant enhanced two-photon ionization through the previously unobserved $(3)^{3}\Pi \leftarrow a^{3}\Sigma^{+}$ band.

\footnotetext{\textit{$^{a}$~Istituto Nazionale di Ottica, INO-CNR, U.O.S. Pisa "Adriano Gozzini", via Moruzzi 1, 56124 Pisa, Italy Fax: 39-0503152522; Tel: 39-0503152529; E-mail: carlo.gabbanini@ino.it}}
\footnotetext{\textit{$^{b}$~Laboratoire Aim\'{e} Cotton, CNRS, B\^at. 505, Univ Paris-Sud,\\F-91405 Orsay Cedex, France. }}


\footnotetext{}

\section{Experiment}
\label{sec:experiment}

The experiment is done in a double species magneto-optical trap (MOT) loaded from vapor and produced inside a UHV metal chamber. The rubidium vapor
pressure is established by running current through a metal dispenser while for Cs a metal reservoir is used, separated from the main chamber by a valve.
The cooling laser for cesium is a DFB diode laser (150~mW power), with frequency tuned two linewidths below the $6S_{1/2}(F=4)\rightarrow 6P_{3/2}(F=5)$ transition. For $^{85}$Rb the cooling laser is a DFB diode laser (80~mW power), frequency tuned two linewidths below the $5S_{1/2}(F=3)\rightarrow 5P_{3/2}(F=4)$ transition. Two other diode lasers tuned on transitions from the other ground hyperfine levels of the two atoms act as repumpers. The repumper beams pass through acousto-optic modulators and the first diffracted order beams are used for the MOT. In this way the repumper beams can be almost extinguished, with the remaining intensity due to the extintion ratio of the AOM. All lasers are frequency-locked in separated cells through saturated absorption spectroscopy.

The double MOT has separated horizontal arms with retroreflected beams for the two species, while it has superposed beams along the vertical direction, that is also the axis of the quadrupole magnetic field. This configuration allows for independent control of the alignment and therefore for maximizing the overlap of the two cooled samples. The overlap is monitored by two CCD cameras along orthogonal axis. The double species MOT captures a similar number of Cs and $^{85}$Rb atoms (nearly 10$^{7}$) with densities of a few 10$^{10}$~cm$^{-3}$.

The PA step is performed by a tapered amplifier that is injected by a dedicated DFB diode laser. A light power of 0.7 W is available to be focused on the trapped sample. Thanks to its large bandwidth and to robust injection, the tapered amplifier remains injected for all the tuning range of the DFB diode laser, that by changing its temperature exceeds thirty wavenumbers. The absolute PA laser frequency is measured by a wavemeter, while a Fabry-P\'erot interferometer monitors the frequency scan. The FP interferometer can be also used to frequency lock the DFB laser.

\begin{figure}[h]
  \includegraphics[width=0.4\textwidth]{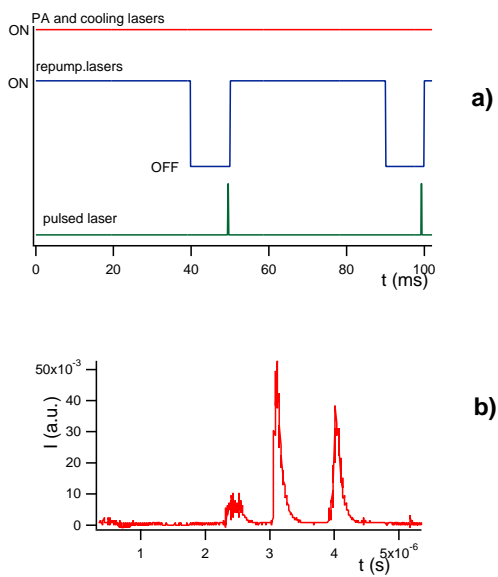}
  \caption{(a) Time sequence of the experiment. (b) an example of a time-of-flight record; the three peaks from left to right correspond to Rb$^+$, Cs$^+$ and RbCs$^+$ ion signals, respectively.}
  \label{fig:fig1}
\end{figure}

The molecules are resonantly ionized by a laser pulse, that is given by a dye laser (Quantel TDL50) pumped by the second harmonics of a Nd:YAG laser (Continuum Surelite I-20, with 20 Hz repetition rate and 7~ns time width). The dye laser operates with LDS 698 dye and covers the frequency range from 13800 to 14800 cm$^{-1}$. The pulsed beam, with  energy of about 1~mJ, is softly focused to the cold sample. The pulsed laser is fired after that the repumping lasers of the double MOT have been almost extinguished for about 10 ms, making temporal dark SPOTs. This has the double advantage of an initial increase of the atomic densities and, even more important, a decrease of the collisional losses (both single species and inter-species~\cite{harris2008}) that are particularly strong with atoms in the excited states. The time sequence of the experiment is shown in Fig.\ref{fig:fig1}a. After the laser pulse the produced atomic and molecular ions are repelled by a grid, separated by time-of-flight and detected by a microchannel plate. The ion signals are recorded by gated integrators, and a typical example is displayed in Fig.\ref{fig:fig1}b.

\section{Experimental results}
\label{sec:results}

\begin{figure}[h]
  \includegraphics[ width=0.45\textwidth]{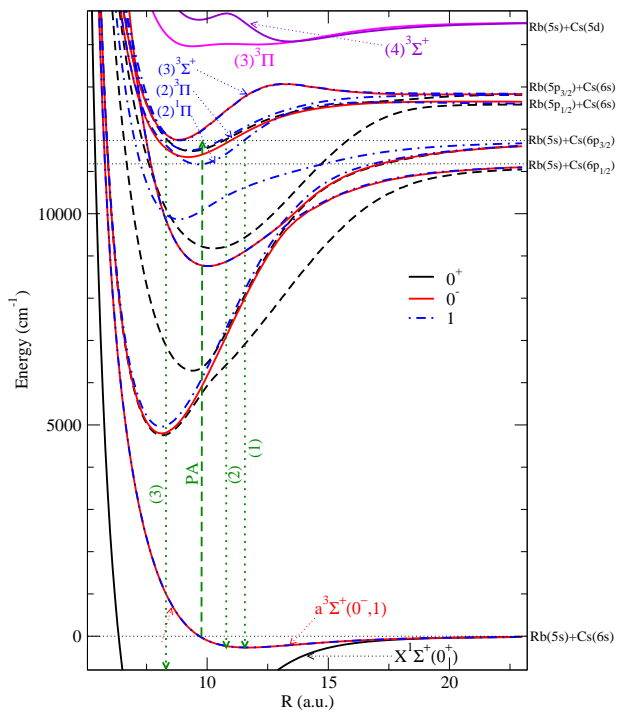}
  \caption{Relevant potential curves \cite{fahs2002} for the RbCs formation process and detection discussed here. Note that states with total electronic angular momentum projection $\Omega=2$ are not displayed, for clarity. Curves are labelled with their Hund's case $c$ symmetry notation. Some  curves correlated to Rb(5$P_{1/2,3/2}$)+Cs($6S$) are also labelled with their Hund's case $a$ symmetry. The upward arrow suggests the probable path for the photoassociation (PA), and numbered downward arrows possible paths for spontaneous emission toward stable RbCs molecules.}
  \label{fig:fig2}
\end{figure}

The RbCs molecule formation has been investigated by scanning the PA laser over all its tuning range below the Rb(5S$_{1/2}$)+Cs(6P$_{3/2}$) dissociation limit (about 15 cm$^{-1}$) and by changing wavelength region of the pulsed laser for detection. As it can be deduced from Fig.\ref{fig:fig2} and as discussed by Kerman et al~\cite{kerman2004}, the PA process below the Rb(5S$_{1/2}$)+Cs(6P$_{3/2}$) dissociation limit has the drawback to induce predissociation to states correlated to the lowest excited asymptote (i.e. Rb(5S$_{1/2}$)+Cs(6P$_{1/2}$)), however this does not preclude to find specific paths that efficiently produce (meta)stable molecules by radiative stabilization. In fact RbCs molecule formation has been observed \textit{only} when the PA laser is tuned -8.1 cm$^{-1}$ below the asymptote and with a REMPI laser wavelength of about 707 nm. The time-of-flight spectrum of Fig.\ref{fig:fig1}b is recorded under such conditions.

\begin{figure}[h]
  \includegraphics[width=0.45\textwidth]{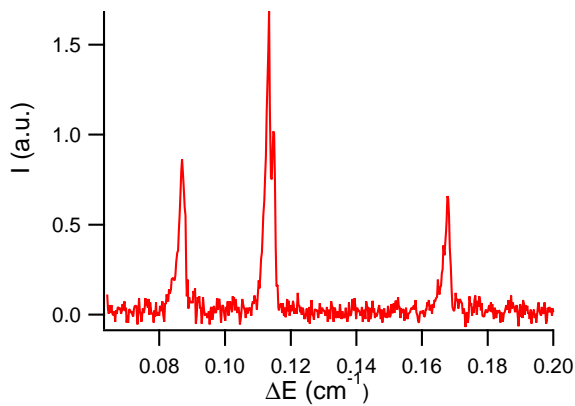}
  \caption{Rotational structure of the PA line at -8.1 cm$^{-1}$. The three peaks correspond to J=0,1,2 rotational levels of the excited molecular state.}
  \label{fig:fig3}
\end{figure}

\begin{figure}[h]
  \includegraphics[width=0.45\textwidth]{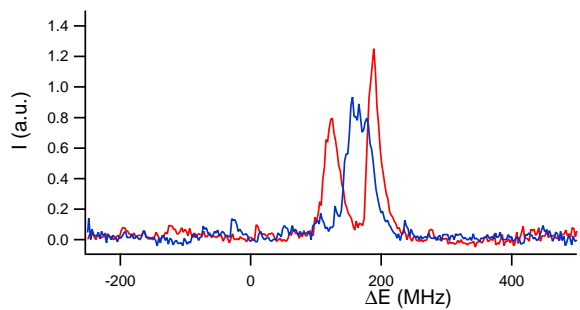}
  \caption{Stark effect on the J=1 line: the line is recorded for two different values of static electric field E$_{el}$ with E$_{el}^{red}$$>$E$_{el}^{blue}$.}
  \label{fig:fig4}
\end{figure}

No other PA lines have been observed over all the scanned region. In particular no line with small B$_v$ which could belong to the Rb(5S$_{1/2}$)+Cs(6P$_{3/2}$) asymptote has been observed, in accordance with the hypothesis of a strong predissociation. 
The PA line has a simple rotational structure that is plotted in Fig.\ref{fig:fig3}. The three peaks are due to the $J=0,1,2$ rotational levels of the excited molecular state and correspond to a rotational constant equal to 0.0138(7)~cm$^{-1}$. The $J=1$ line feels the Stark effect due to the static electric field used to extract the ions from the MOT region and direct them to the microchannel plate. The static electric field can be varied from approximately zero, by applying a pulsed electric field to the grid after the laser pulse, to values in the hundreds V/cm range. Two recordings of this line using different electric fields for extraction are shown in Fig.\ref{fig:fig4}. In the present setup the electric field at the MOT position is not precisely known, however the electric dipole moment of the excited molecular state can be estimated between 3 and 4 Debye.

\begin{figure}[h]
  \includegraphics[width=0.45\textwidth]{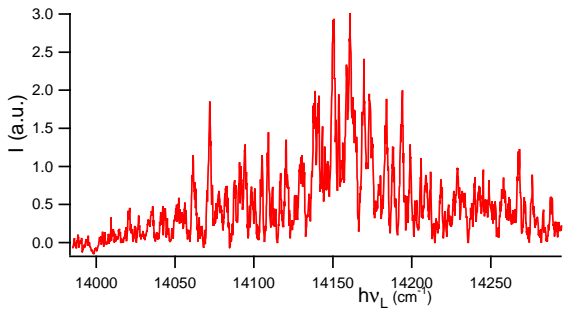}
  \caption{Spectrum of RbCs$^+$ molecular ions recorded by scanning the pulsed dye laser in the 700-715~nm wavelength range. The PA laser is locked on the $J=1$ line at 852.946~nm.}
  \label{fig:fig5}
\end{figure}

By locking the PA laser to a stable Fabry-P\'erot interferometer it is possible to probe the molecular sample by scanning the detection laser. A spectrum of the detected RbCs$^+$ ions with the PA laser locked at 852.946~nm ($J=1$ line) is shown in Fig.\ref{fig:fig5}. The spectrum extends in the 700-715~nm wavelength range and has a complex structure. The observations are consistent with the explanation that triplet ground state molecules are produced and subsequently ionized by REMPI through the previously unobserved $(3)^{3}\Pi \leftarrow a^{3}\Sigma^{+}$ band. The $(3)^{3}\Pi$ state has a local minimum and it is correlated with the Rb($5^2S$)+Cs($5^2D$) dissociation limit~\cite{allouche2000}.

\begin{figure}[h]
  \includegraphics[width=0.45\textwidth]{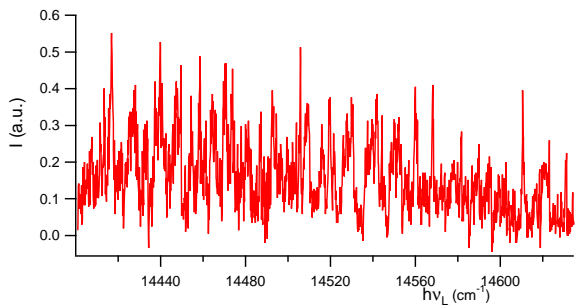}
  \caption{Spectrum of RbCs$^+$ molecular ions recorded by scanning the pulsed dye laser in the 683-694~nm wavelength range. The PA laser is locked on the $J=1$ line at 852.946~nm.}
  \label{fig:fig6}
\end{figure}

As it can be deduced from Fig.\ref{fig:fig2}, the bound states of the $a^{3}\Sigma^{+}$ state oscillate in the region r$\geq$9.7 $a_0$, while in order to match the RbCs$^+$ potential curve, the ionization should occur at r$\leq$13 $a_0$. The Franck-Condon factors for the $(3)^{3}\Pi \leftarrow a^{3}\Sigma^{+}$ band have been calculated to be particularly favorable~\cite{beuc2007}. Another weaker detection region has been observed in the 683-693~nm wavelength range (Fig.\ref{fig:fig6}). It is tentatively assigned to the $(4)^{3}\Sigma^{+} \leftarrow a^{3}\Sigma^{+}$ band. As shown in Fig.\ref{fig:fig2}, the $(4)^{3}\Sigma^{+}$ is also a double well state correlated with the Rb($5^2S$)+Cs($5^2D$) dissociation limit, but its local minimum is above that limit, implying that the supported bound states have finite lifetimes. The interpretation of the molecular bands, that will be the subject of a future study, will allow to determine the vibrational distribution of the produced RbCs molecules. Similar results, at different detuning of the PA laser, are obtained for $^{87}$RbCs molecules, starting from a double species trap of $^{87}$Rb and Cs atoms.

A precise value for the molecular formation rate cannot be given because the ionization and detection efficiencies are not known. However the RbCs$^+$ detection rate is just a factor two lower than the Cs$_{2}^+$ detection rate measured in the same experimental condition photoassociating cold cesium atoms through a giant line~\cite{fioretti1998,bouloufa2010} and detecting by REMPI through the $(2)^{3}\Pi_g \leftarrow a^{3}\Sigma_{u}^{+}$ band. If a similar ionization efficiency is assumed, the RbCs formation rate can be estimated to be above 10$^4$~molecules/s.

\section{Interpretation: photoassociation of RbCs at short-range}
\label{sec:model}

The most striking feature of the present observation is the large rotational constant of the detected line, and the absence of neighboring lines. This strongly suggests that it corresponds to the direct population of a vibrational level lying at the bottom of a molecular potential well located in the energy range of the Rb(5S$_{1/2}$)+Cs(6P$_{3/2}$) dissociation limit. Fig.\ref{fig:fig2} illustrates the probable process using the potential curves including molecular spin-orbit computed in Ref.\cite{fahs2002}. Indeed, several potential wells correlated to the Rb(5P$_{1/2}$)+Cs(6S$_{1/2}$) dissociation limit have their minimum located just below Rb(5S$_{1/2}$)+Cs(6P$_{3/2}$): an isolated $\Omega=1$ state (in Hund's case $c$ notation with $\Omega$ being the projection of the total electronic angular momentum on the molecular axis) mostly induced by a $(2)^1\Pi$ state, and a group of four states ($\Omega=0^+, 0^-, 1, 2$) resulting from the spin-orbit splitting of the $(2)^3\Pi$ state. In the following, these states are numbered by increasing energy  $(4)1$, $(4) 0^{-}$, $(5) 0^{+}$, $(5)1$, $(2)2$, respectively.

The observed structure starts with a $J=0$ line and does not exhibit a detectable hyperfine structure, and is therefore assigned to one of the two $(4) 0^{-}$, $(5) 0^{+}$ molecular states above. The measured rotational constant is in good agreement with the ones computed in Ref.\cite{fahs2002} (0.0136 ~cm$^{-1}$ and 0.0133~cm$^{-1}$, respectively). The lower levels of these potentials can be populated by the PA laser in the vicinity of the inner turning point of the lowest triplet  $a^{3}\Sigma^{+}$ potential, as illustrated by the arrow marked "PA" in Fig.\ref{fig:fig2}. In contrast with what is generally reported (see for instance Ref.\cite{jones2006}), this result demonstrates the possibility to achieve PA of a pair of different ultracold atoms at short distances. There is only one another observation of this kind reported up to now \cite{deiglmayr2008a}: the lowest levels of the $(1)^1\Pi$ state correlated to the Li($2^2S$)+Cs($6^2P$) dissociation limit of the LiCs molecule have been successfully populated by PA from the inner turning point of the $a^{3}\Sigma^{+}$ potential as well. The apparent violation of the spin selection rule is explained by the small spin-orbit mixing of the $(1)^1\Pi$ with triplet states \cite{deiglmayr2009}, also expected in RbCs.

Once such excited levels are populated, they naturally decay by spontaneous emission down to a broad range of $a^{3}\Sigma^{+}$ vibrational levels as pictured by arrow (2) in Fig.\ref{fig:fig2} starting from levels of the $(2)^3\Pi$ manifold. As the spin-orbit is large in the RbCs molecule, another possible path (arrow (1) in Fig.\ref{fig:fig2}) could be the decay from the $(2)^1\Pi$ state (or the (4)1 state in Hund's case $c$) as it is likely coupled to neighboring triplet states just like in LiCs. The precise distribution of the populated vibrational levels will be analyzed from the spectra reproduced in Figs.\ref{fig:fig5} and \ref{fig:fig6}. It is also likely that some of the excited levels decay down to the lowest levels of the $X^{1}\Sigma^{+}$ ground state (see arrow (3) in Fig.\ref{fig:fig2}). This should be analyzed using another detection scheme where excited singlet states would be reached by the first photon of the REMPI process.

\section{Conclusion}
The formation of translationally and rotationally cold RbCs molecules in a double species MOT is observed, and assigned to the short-range photoassociation of ultracold pair of atoms followed by radiative stabilization to the lowest metastable state, $a^{3}\Sigma^{+}$. They are detected by resonant enhanced two-photon ionization through two different paths including the previously unobserved $(3)^{3}\Pi \leftarrow a^{3}\Sigma^{+}$ band. Further experiments should explore the vibrational distribution of the produced ultracold molecules, as well as the possibility to create ground state molecules through other isolated PA lines which are expected in the same energy range. In particular detunings below the -8~cm$^{-1}$ region investigated here should be explored, as several other levels are expected below the energy of the Rb(5S$_{1/2}$)+Cs(6P$_{3/2}$) limit. This would provide the precise location of the $v=0$ levels of the $(4)1$, $(4) 0^{-}$, $(5) 0^{+}$, $(5)1$, $(2)2$ potentials, to be compared with theoretical calculations. Blue-detuned photoassociation (\textit{i.e.} at positive detunings of the PA laser above Rb(5S$_{1/2}$)+Cs(6P$_{3/2}$)) should be possible as well, as recently observed by Bellos \textit{et al.} in the same issue of PCCP. Such simple formation path of RbCs molecules can be useful as a first step for a further optical manipulation that modifies the internal degrees of freedom to produce a sample of ultracold polar molecules.

\section*{Acknowledgments}
The author wishes to thank A.Fioretti for helpful discussions, D.Comparat for lending some experimental equipment and M.Voliani and F.Pardini for technical support.




\footnotesize{
\bibliography{rbcs} 
\bibliographystyle{rsc} 
}

\end{document}